# Effect of temperature and excitation power on down-conversion process in Tb$^{3+}$/Yb$^{3+}$-activated silica-hafnia glass-ceramic films

Salima El Amrani [a,*], Michael Sun [b], Sirona Valdueza-Felip [b], Fernando B. Naranjo [b], Mohammed Reda Britel [a], Maurizio Ferrari [c], Adel Bouajaj [a]

[a]Laboratory of Innovative Technologies, National School of Applied Sciences of Tangier, Abdelmalek Essaadi University, Tetouan, Morocco
[b] Photonic Engineering Group, Electronic Department, University of Alcala, Madrid-Barcelona Road km 33.6, 28871, Alcala de Henares, Spain.
[c] IFN-CNR CSMFO Lab and FBK Photonics Unit, Via Alla Cascata 56/C Povo, 38123, Trento, Italy.

*salima.elamrani1@etu.uae.ac.ma

## Abstract

Transparent glass ceramics, when activated by rare earth ions, are excellent photonic materials. Regarding photonic glass-ceramics based on silicates, hafnia and silica in a binary system has proved to be an excellent matrix to incorporate rare earth ions in the hafnia nanocrystals, resulting in important luminescence enhancement and, consequently, allowing a large spectrum of critical applications. Here we will focus on the down-conversion mechanism driven by the couple Tb$^{3+}$/Yb$^{3+}$, largely exploited in photovoltaic systems. The research presented here has been performed on 70SiO –30HfO silica-hafnia glass-ceramic films activated with 19 % rare earth ions: [Tb + Yb]/[Si + Hf] = 19 %. Two main results will be discussed: (a) the intensity and the broadening of the Yb$^{3+}$ emission band at 975 nm were found to be temperature-dependent, as shown in the figure; (b) the energy transfer mechanism Tb$^{3+}$→Yb$^{3+}$ will be discussed, referring to the mechanisms that have been proposed in the literature. In relation to the latter topic, the power dependence spectra for the luminescence of Yb$^{3+}$:$^2$F$_{5/2}$ → $^2$F$_{7/2}$ will be discussed.

# 1. Introduction

Renewable energy is on the rise across the globe, with solar energy at the forefront of the green energy revolution. Indeed, solar cell efficiency must be improved in order to satisfy future energy demands in the coming years.

Shockley-Queisser limit denotes the maximum energy efficiency attainable with solar cells [1]. The energy losses limit the conversion efficiency of crystalline silicon PV cells to 30 percent [1,2]. Photons with energies below the bandgap are unable to produce electron-hole pairs, and their energy is completely wasted. Moreover, high-energy radiation is inefficient since most of its energy is lost as heat [3].

Therefore, the key to exceeding the limit of Shockley-Queisser and obtaining extremely effective solar energy conversion is to better align the semiconductor's band gap with the sun's spectrum [4]. Thus, two spectrum modification strategies could achieve this: adapt the device's band gap of the sun's spectrum (as in solar cells with multiple junctions); or modify the light wavelengths reaching the solar cell (down-conversion) [5]. We adopted the second approach, which consists of dividing a photon having a wavelength slightly less than 500 nm using the down-conversion technique to produce two photons at about 1000 nm that can easily reach the band gap of c-Si [6].

Due to their rich electronic level distribution in the visible and infrared, rare-earth ions are one of the most commonly used down-converting materials [7]. The $Tb^{3+}/Yb^{3+}$ rare earth ions were selected for this study due to their interaction transfer energy, which can permit the appearance of a quantum cutting process. The $Tb^{3+}$: $^5D_4$ energy level corresponds to almost double the energy of the $Yb^{3+}$: $^2F_{5/2}$ energy level. $Tb^{3+}$, acting as a sensitizer, absorbs a photon in the blue at 488 nm through the $^7F_6 \rightarrow {}^5D_4$ energy transition and excites two $Yb^{3+}$ ions [8]. Two near-infrared photons are thus generated by $Yb^{3+}$ ions through the $^2F_{5/2} \rightarrow {}^2F_{7/2}$ following a single photon absorption by a $Tb^{3+}$ ion (Fig. 1).

Regarding the choice of the matrix, $SiO_2$-$HfO_2$ is a promising material for producing luminescent glass ceramic layers for rare earth doping, due to their remarkable mechanical and optical characteristics. Besides, it has been proven that the cut-off frequency of the rare earth ions encased in hafnia nanocrystals is around 700 cm$^{-1}$ [9].

As a result, the introduction of hafnia nanocrystals leads to an elongation of the measured emission lifetime and a reduction in non-radiative transition processes [10]. Additionally, glass-ceramic materials have the potential to surpass the performance of glasses through combination of glass's beneficial characteristics with the improved spectroscopic properties of crystals [10,11].

Among several techniques of thin film fabrication, the sol-gel route is a less expensive technique relevant to the synthesis of silica-hafnia luminescent layers with good optical quality [12,13]. At this aim, a silica-hafnia glass-ceramic synthesized via the sol-gel technique constitutes an appropriate matrix for the production of rare-earth luminescent layers for down conversion. The down-converter layer sits in front of the photovoltaic cell, modifying the light before it reaches the solar cell.

Studies previously done on $70SiO_2$–$30HfO_2$ down-converting luminescent layers co-doped with $Tb^{3+}$ and $Yb^{3+}$ ions revealed that, [Yb]/ [Tb] = 4 defines the optimal ratio of dopant concentration between the two. This has been done by gradually increasing the quantity of $Yb^{3+}$ ions (acceptors) in accordance with a certain amount of $Tb^{3+}$ ions (donors) until reaching the optimal ratio, which after this, the increase of the acceptor can be detrimental [14]. Furthermore, the transfer efficiency of glass-ceramics is higher than that of glasses, and it rises with the overall rare earth concentration [15]. Thus, the highest efficiency (190 %) was found with 70SiO2–30HfO2 luminescent down-converters co-doped with 19 % of Tb3+/Yb3+ ions [9].

The goal of the research is to assess the down-conversion and luminescence efficiencies as a function of the operating temperature and the excitation power of 70SiO2–30HfO2 films, activated by 19 % molar concentrations of rare earths [Tb + Yb]/[Si + Hf] = 19 %, synthesized by the sol-gel route using the spin-coating technique. All samples consisted of glass-ceramic (GC) treated up to 1000 ◦C. XRD, transmission, and photoluminescence characterization results are reported.

## 2. Experimental

Tb3+/Yb3+ co-doped 70SiO2–30HfO2 planar waveguides are obtained by the sol-gel route using the spin-coating technique, following the synthesis protocol described in Ref. [16]. On well cleaned pure SiO2 substrates, silica-hafnia films were deposited with a spinning rate of 2500 rpm for 30 s. After 20 depositions, the resulting films were heated in the air at 900◦ for 5 min to make glass samples. Glass-ceramic (GC) samples were produced by nucleating hafnia nanocrystals inside the film through the use of an additional heat treatment at 1000 ◦C in air for 30 min. Thus, rare earth ion-doped 70SiO2–30HfO2 GC films were produced. In this study, the following characterization techniques were employed: transmission, photoluminescence, and XRD.

X-ray Diffraction (XRD) spectra were collected in PSD Fast Scan mode in the 2θ range 5–90°, with a scanning step of 0.06°.

All transmittance tests were performed between 350 and 1700 nm using a UV near-infrared spectrophotometer.

Photoluminescence (PL) measurements were performed using a diode emitting continuous-wave, which was focused into a 1-mm-diameter spot. The emission of a 193-mm-focal-length Andor spectrograph with a 70 charge-coupled device camera that operates at -60 °C and is equipped with a UV-extended silicon-based camera. In addition, in order to eliminate the laser effect, wavelengths below 495 nm are filtered with a 495 nm filter. In order to acquire measurements at low temperatures, samples were placed in a cryogenic system that reaches 11 K.

## 3. Results and discussion

### 3.1. X-ray diffraction analysis

X-ray diffraction analysis was used to characterize the structural parameters of the produced waveguides. Fig. 2 presents the XRD patterns of 70SiO2–30HfO2 glass-ceramic waveguides after heat treatment at 1000 °C doped with a total molar concentration of 19 % of $Tb^{3+}/Yb^{3+}$ ions with and without ytterbium.

XRD spectra show that the films contain an amorphous structure represented by a clear hump centered at 2θ ≈ 21°, derived from the luminous layers' SiO2 composition (70 mol%) and the silica substrate.

Following 30 min of heat treatment at 1000 °C, the 70SiO2–30HfO2 films crystallize, resulting in the presence of Bragg reflection peaks. Some diffraction peaks associated to HfO2 nanocrystals are evident. All of this is consistent with what was reported in Ref. [16].

The [$Tb^{3+}/Yb^{3+}$] co-doped rare earth ions luminescent layers concentration are much higher than that of the single-doped Tb. Thus, by increasing the concentration, the diffraction peaks become more intense, implying an increase in the crystallization effect. This considers how rare earth ions within silica-hafnia glass ceramic films are incorporated into hafnia nanocrystals. Gonçalves et al. found comparable results in the $Er^{3+}$ planar waveguides doped with 70SiO2–30HfO2 [17].

### 3.2. Transmittance analysis

Fig. 3 displays the optical transmission spectra of $Tb^{3+}$-activated 70SiO2–30HfO2 glass ceramic films with and without $Yb^{3+}$ ions.

Optical spectra demonstrate interference fringes caused by the disparity in refractive index between the v-SiO2 slabs and the glass-ceramic layers [18]. The visible infrared transmittance of 70SiO2–30HfO2 glass ceramic layers doped with 19 % Tb3+ with (b) and without (a) Yb3+ is on the order of 90 %. This high value is a result of the precipitated crystals' tiny size, which is smaller than the visible wavelength. Identical results were reported in Ref. [16]. On the other hand, the transmittance of the co-doped glass-ceramic samples (b) is slightly lower than the single-doped ones (a). The increase of the light scattering should be attributed to the highest content of nanocrystals induced by the Yb3+ addition as suggested by the XRD spectra.

*3.3. Photoluminescence study*

*3.3.1. Temperature-dependent photoluminescence*

Fig. 4 shows the temperature dependence of the photoluminescence spectra in the near-IR region for the GC sample under 488 nm excitation. The PL peaks position does not show a discernible change during the temperature variation from 11 K up to 300 K. For the whole temperature range, the energy corresponding to the maximum of the PL band exhibits a shift well described by the temperature dependent population processes. The intensity of the PL emission significantly depends on the temperature due to the competition between the efficiency of non-radiative and radiative relaxation processes. Indeed, as the temperature increases, phonon-assisted relaxation and energy transfer process, where energy is either transformed into heat (multiphonon relaxation) or transmitted to another ion (energy transfer), leads to 976 nm PL weak intensities. As the temperature decreases, the contribution of nonradiative recombination falls while that of radiative transitions increase [19].

The FWHM of the PL spectrum of the 2F5/2 → 2F7/2 transition also displays a dependence on temperature, with the linewidth substantially broadening from 8.7 to 11.3 nm as the temperature rises from 11 to 300 K, as shown in Fig. 5.

To determine the origin of this broadening, the full width of the emission line is deconvoluted using Voigt profiles, resulting in homogeneous and inhomogeneous broadening [20].

Fig. 6 shows the homogeneous and inhomogeneous broadening of the 2F5/2 → 2F7/2 transition as a function of temperature. The inhomogeneous broadening occurs due to site variation caused by a random distribution of local crystal fields leading to a Gaussian line shape, a Lorentzian line shape is the result of homogeneous broadening caused by dynamic perturbations on energy levels that affect all ions equally.

The inhomogeneous broadening doesn't show a discernible temperature dependence as expected; meanwhile, homogeneous widths exhibit a simple power law dependence on temperature over the entire temperature range, and this can be explained by the fact that high temperatures have the potential to increase the non-radiative energy transfer mechanisms as was shown in the previous section, thereby amplifying the probability of energy transfer among the dopant ions. The augmentation of energy transfer can lead to a wider spectral line and, thus, an increase in homogeneous widths.

Therefore, controlling temperature is crucial in the process of down-converting $Tb^{3+}/Yb^{3+}$ co-doped glass ceramic materials. This is necessary in order to limit the FWHM broadening and optimize the efficiency of the down conversion.

*3.3.2. Excitation power-dependent photoluminescence*

Fig. 7 (a) shows the photoluminescence emission spectra of the glass ceramic sample co-doped with 19 % $Tb^{3+}$ and $Yb^{3+}$ obtained under 476 nm excitation. The spectra exhibit the typical $2F_{5/2} \rightarrow 2F_{7/2}$ transitions of $Yb^{3+}$ ions. The emission of the $Yb^{3+}$ ions after excitation in the blue region demonstrates an effective energy transfer from $Tb^{3+}$ to $Yb^{3+}$ and, thus, an efficient down-conversion process.

On the other hand, a more comprehensive understanding of the energy transfer mechanism can be attained through an examination of the power dependence characteristics of the $Yb^{3+}$ 980 nm near-infrared emission.

The photoluminescence (PL) spectra of Ytterbium ($Yb^{3+}$) ions undergoing a $2F_{5/2} \rightarrow 2F_{7/2}$ transition for varying excitation powers are recorded at 11 K. Thus, increasing the laser power from 25 to 270 mW has no effect on the PL peak position, and the maxima maintain the same wavelength position at 978 nm. In addition, as the pump intensity increases, more blue photons are absorbed by the $Tb^{3+}$ ions, leading to a higher population of excited $Yb^{3+}$ ions and an increase in the emission intensity at 980 nm.

While increasing laser power, the intensity keeps increasing, even at high pumping powers. This linear behavior indicates that we are far from saturation effects. Fig. 7(b).

In addition, the correlation between the integrated PL intensity and the pumping laser power at 980 nm can yield valuable information regarding the number of photons engaged during the excitation process.

The subsequent equation expresses the relationship between the PL intensity of down-conversion ($I_{DC}$) and the intensity of the pumping laser ($I_P$).

$$I_{DC} = \alpha (I_P)^n \quad (1)$$

Where $\alpha$ denotes a factor of proportionality, while the exponent n signifies the quantity of photons that participate in the process of down-conversion [21,22].

The number of photons n calculated from the linear-fitted line's slope coefficient is 1.

So, during the quantum cutting process, approximately one blue photon converted two near-IR photons. So, these results confirm the cooperative energy transfer mechanism in Tb3+/Yb3+ down-converting glass-ceramic material studies in this work, where two Yb3+ ions are emitted through the 2 F5/2 → 2 F7/2 transition following the excitation of one single Tb3+ ion to the 5D4 energy level. More details are reported in Ref. [23].

## 4. Conclusion

70SiO2–30HfO2 glass ceramic films co-doped with Tb3+/Yb3+ with a 19 % molar concentration of rare earths [Tb + Yb]/[Si + Hf] = 19 % were prepared by the sol-gel method using spin coating processing. The samples have 90 % visible infrared transmittance, making them useful for PV applications. The XRD patterns show crystallization after 30 min at 1000 ◦C. The photoluminescence spectra revealed that the sample exhibited temperature-dependent peak intensity and full width at half maximum (FWHM), suggesting that lower temperatures may reduce non-radiative relaxation processes allowing enhancement of the emission intensities of the rare earth ions. Finally, the emission dependence on excitation power validated the cooperative energy transfer mechanism in the Tb3+/Yb3+ down-converting glass-ceramic material. These results confirm that SiO2- HfO2 glass ceramic films, doped with Tb3+/Yb3+ rare earth ions, have the potential to serve as efficient down conversion filters for photovoltaic applications.

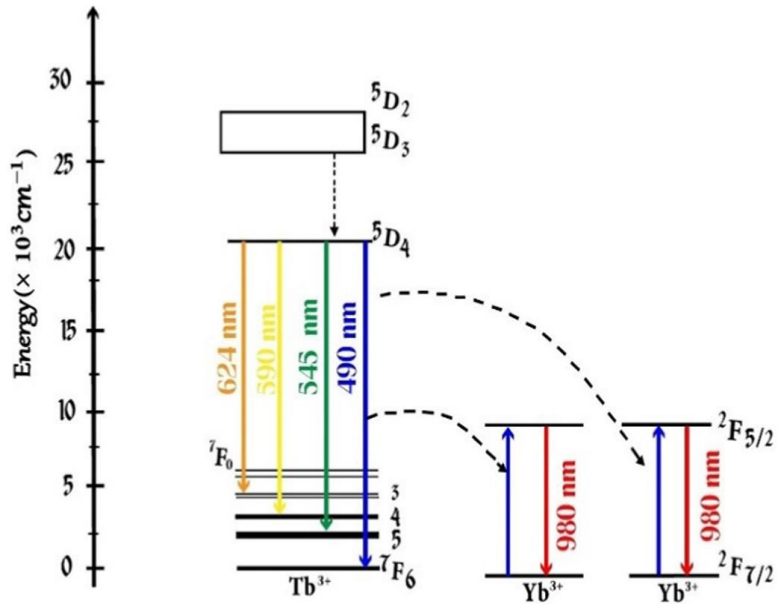

**Fig. 1.** Quantum cutting (QC) mechanism simplification between Tb3+ and Yb3+.

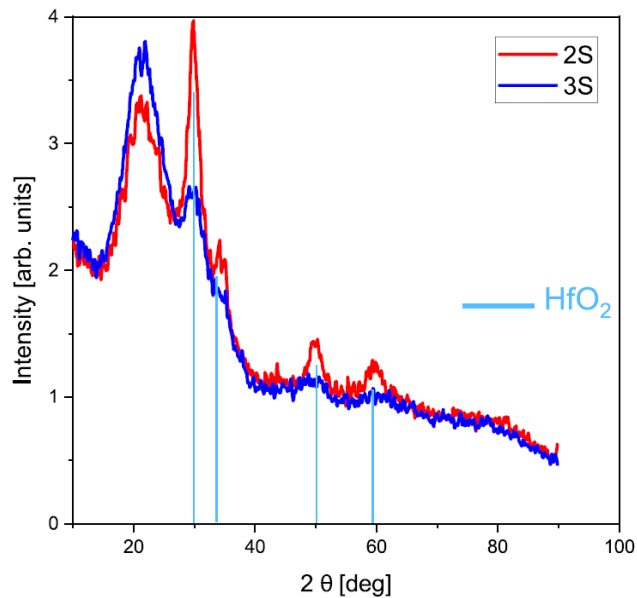

**Fig. 2.** XRD spectra of Tb3+-activated 70SiO2–30HfO2 glass ceramic films with (2S) and without (3S) Yb3+ ions.

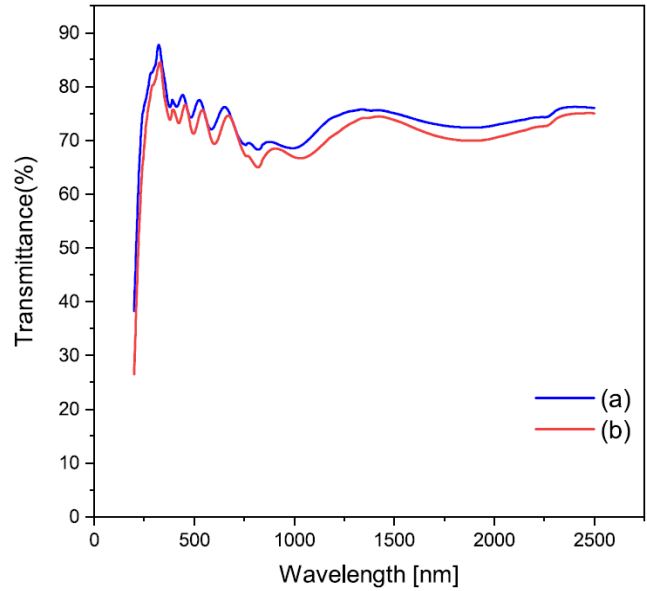

**Fig. 3.** Optical transmission spectra of Tb3+-activated 70SiO2–30HfO2 glass ceramic films with (b) and without Yb3+ions (a).

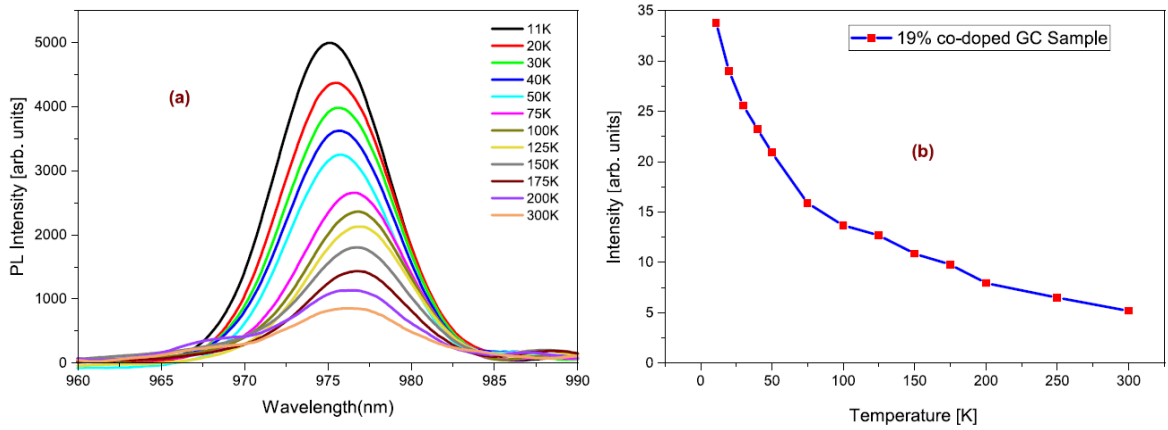

**Fig. 4.** (a) PL emission spectra in the near IR region for the 19 % co-doped GC sample under 488 nm excitation as function of temperature; (b) Peak intensity as function of temperature.

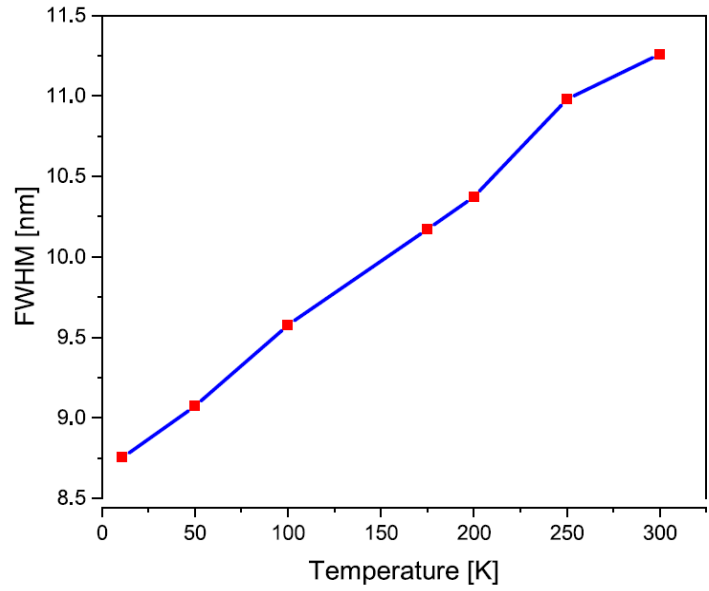

**Fig. 5.** FWHM of the emission spectra as function a of temperature.

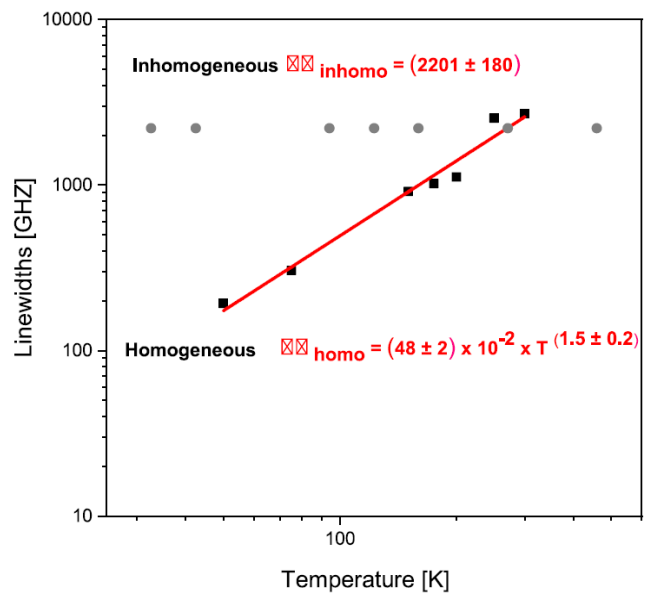

**Fig. 6.** Homogeneous and inhomogeneous broadening as a function of temperature.

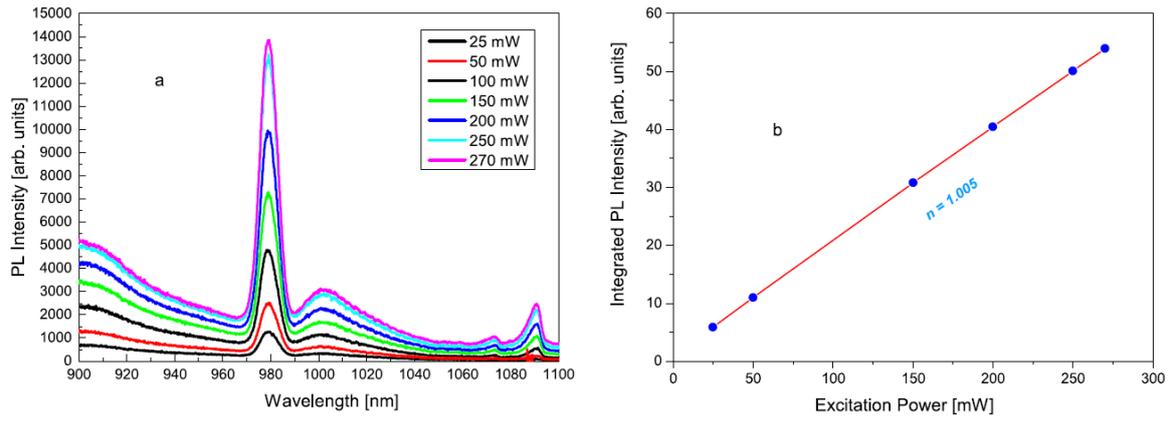

**Fig. 7.** 2 F5/2 → 2 F7/2 PL emission spectra at 11 K, for the 19 % co-doped GC sample, under 488 nm excitation, with different excitation power (a); the linear dependence assures for non-saturation conditions (b).